# Effect of Si on Fe-rich intermetallic formation and mechanical properties of heat-treated Al-Cu-Mn-Fe alloys

Yuliang Zhao[1,2], Weiwen Zhang[1]*, Chao Yang[1], Datong Zhang[1], Zhi Wang[1]

[1]National Engineering Research Center of Near-net-shape Forming for Metallic Materials, South China University of Technology, Guangzhou, 510641, China

[2]School of Engineering & Computer Science, University of Hull, East Yorkshire, HU6 7RX, UK

*Corresponding author. Tel: +86-20-87112022, Fax: +86-20-87112111.

E-mail: mewzhang@scut.edu.cn

**Abstract**: The effect of Si on Fe-rich intermetallics formation and mechanical properties of heat-treated squeeze cast Al-5.0Cu-0.6Mn-0.7Fe alloy was investigated. Our results show that increasing Si content promotes the formation of $Al_{15}(FeMn)_3(SiCu)_2$ (α-Fe), and varying the morphology of T ($Al_{20}Cu_3Mn_2$) where the size decreases and the amount increases. The major reason is that Si promotes heterogeneous nucleation of the intermetallics leading to finer precipitates. Si addition significantly enhances ultimate tensile strength and yield strength of the alloys. The strengthening effect is mainly owing to the dispersoid strengthening by increasing volume fraction of T phase and less harmful α-Fe with a compact structure, which make the cracks more difficult to initiate and propagation during tensile test. The squeeze cast Al-5.0Cu-0.6Mn-0.7Fe alloy with 1.1% Si shows significantly improved mechanical properties than the alloy without Si addition, which has tensile strength of 386 MPa, yield strength of 280 MPa and elongation of 8.6%.

## I. Introduction

Al-Cu alloys have been widely used in automobile manufacturing, space technology and aerospace industry owing to their high specific strength, good heat resistance and excellent fatigue properties [1-3]. To meet both recyclable use and increasing demands of Al alloys, recycling Al alloys have become an important source of Al production [4-6]. For example, the consumption of recovered aluminum in US in 2015 was ~ 3.61 million ton which is about 46% of the production came from old aluminum scrap [7]. Moreover, recycling Al alloys production creates only approximately 4% as much $CO_2$ as by primary production.

As a typical high-strength Al-Cu alloy, Al-5.0Cu-0.6Mn is wide applications in room and elevated temperatures because of its excellent mechanical properties [8]. However, one of the greatest challenges to aluminum recycling is the accumulation of impurities elements, i.e., Fe, Si, Ni, Zn, Mg and Mn, in the recycling Al alloys, which can cause sharply degradation on the mechanical properties (ductility, formability and fatigue properties) [9-10]. Fe is the most common impurity elements in the Al-Cu scraps and it is difficult to eliminate [11, 12]. For high performance Al-Cu alloys, Fe and Si contents is usually limited to 0.15 and 0.10 wt. % (hereinafter weight percentage simply as %), respectively [6]. Hence, the relatively tolerant limits pose great challenges for direct reuse of these alloys. Since the solid solubility of Fe in Al-Cu casting alloys is limited, Fe atoms usually precipitate in the forms of hard and brittle Fe-rich intermetallics, such as, Chinese script $Al_{15}(FeMn)_3(SiCu)_2$ (α-Fe) [13-20], $Al_6(FeMn)$ [19-20], $Al_m(FeMn)$ [18, 20] and plate-like $Al_3(FeMn)$ [18-19] and $Al_7Cu_2Fe$ (β-Fe) [15, 19, 20], depending on the alloy composition and cooling rate.

Mn is the most common element added to in Al-Cu cast alloys to minimize their harmful influence on the mechanical properties because Mn can transfer the Fe-rich intermetallics from platelet into Chinese script [14, 18-21]. It was found that the best Mn/Fe mass ratio is 1.6 (without applied pressure) and 1.2 (75 MPa applied pressure) respectively, for completely converted the needle-like β-Fe phase into the Chinese script Fe-rich intermetallic phases [31]. It also was reported that Mn addition promotes the transformation of α-Fe, and their transformation efficient depend on different Fe and Mn content and cooling rate [18-19]. It also contributes to the strength of the alloy through solid solution strengthening [20]. Usually, 0.4-1.0 wt. % Mn is added to the Al-Cu alloys for compensating the negative effect of Fe [10]. Hence, 0.6 % Mn was added into the alloy in the present study.

It is found that the Si/Fe mass ratio has a significantly effect on the Fe-rich intermetallics formation [13-19]. Si can dissolves in the α-Al matrix, excess Si mainly precipitates in the form of Si-containing intermetallics. It has been reported that the additions of ~ 0.1 wt.% Fe and Si resulting in the solid solution strengthening and impurity hardening of purity aluminum alloys [11]. It is found that addition of combined Mn and Si have a higher transformation efficient of β-Fe to α-Fe than individual added [14]. It is observed that B206 alloys obtained the best mechanical properties when the Si/Fe mass ratio is closed to 1 with lower concentration of Fe and Si contents [16]. The modification of Fe-rich intermetallic formation can change the mechanical properties where the α-Fe intermetallic shows less harmful effect [13-16]. Minor addition of Si modified the dispersion, morphology and crystal structure of precipitates of the Al-4Cu-1.3Mg alloy and an associated increase in tensile strengths [17]. The addition of Si in Al-Mg and Al-Mn alloys helps to transform the Fe-rich intermetallics from $Al_6(FeMn)$ to α-Fe [22-24]. The addition of Si to A201 alloy increased the large particles precipitation at grain boundaries and associated enhance of microhardness of the alloy [24]. However, with the increase of Si content, the tensile strength and elongation of T7651 heat-treated 7050 alloys are decreased [25]. It also found that Si addition into the Al-Cu catalyzed the precipitation of θ' phases during aging process [26]. However, the reports on the best Si additive amount in the alloys are conflicting. Thus, the underlying mechanism still needs to be further investigated.

Heat-treatment is one of methods to strengthening Al-Cu alloys with the advantage of dissolution of non-equilibrium phases, elimination of segregation, formation of high density fine precipitates. Al-Cu alloys, also called 2XXX series alloy, are one of the most important precipitation-strengthened alloy systems, because it forms age hardening during the heat treatment process [3]. The $Al_{20}Cu_2Mn_3$ (T) phase usually forms within α-Al matrix of Al-Cu-Mn alloy after solution treatment and aging, which enhance the high temperature deformation resistance of the matrix. Except for the precipitates, the Fe-rich intermetallic phases also experience fragmentation and transformed into different Fe-rich intermetallic phases during heat treatment. According to the previous studies [13, 16], α(CuFe) and α-Fe are the two typical Fe-rich intermetallic phases in the heat-treated Al-Cu alloys. α-Fe is usually formed in the high Si-containing Al-Cu alloys. Because Si is the only element required for solid solution transformation: Si + Al$_6$(FeMn) → α-Al + α-Fe (called 6-α transformation) [27]. This reaction need intake the Si atom from the α-Al matrix and α-Al from the Fe-rich intermetallic phases.

Squeeze casting combines the feature of gravity casting and plastic processing, which can decrease the casting defects and improve the casting quality [28]. Several researchers have reported related studies on squeeze casting of Al alloys [20, 29-34] and got satisfied results. Previous study [20] shows that the elongation of the Al-Cu-Mn-Fe alloys of 75 MPa applied pressure is two times higher than those of 0 MPa alloys. The optimum squeeze casting parameter 2017A wrought Al alloys were revealed: squeeze pressure equal to 90 MPa, melt temperature equal to 700 °C, and die preheating temperature equal to 200 °C [29] . The semisolid slurry of wrought 5052Al alloy and AlCu5MnTi alloy were prepared by the indirect ultrasonic vibration, and then shaped by direct squeeze casting [30, 31]. They found that average diameters of the primary α-Al particles decreased with the increase of squeeze pressure and the tensile properties of the alloy increased. A new technology of near liquidus squeeze casting, can form globular structure without preparation of semi-solid slurries or billets at near liquidus pouring temperature [32]. The squeeze casting techniques can also been using in the prepared of Al based composites alloys [33, 34]. Stir followed by squeeze casting techniques were used to produce A359 composites containing different weight percentage of (SiC + Si$_3$N$_4$) particles [33]. Microstructures of the composites showed a homogeneous and even distribution of hybrid reinforcements within the matrix. The squeeze-cast (SiCp + Ti)/ 7075 Al hybrid composites has been successfully produced [34] and the tensile strength of both composites was improved because of the precipitation hardening of the matrix alloy. Hence, squeeze casting is an attractive and promising technology for producing Al-Cu alloy components with improved mechanical properties.

Up to now, the research on the microstructure evolution and mechanical properties of squeeze cast Al-Cu alloys with high Fe and Si contents during heat treatment is still limited. In present work, we studied the Fe-rich intermetallic phase and their effect on the mechanical properties of heat-treated Al-5.0Cu-0.6Mn-0.7Fe alloys with different Si addition produced by squeeze casting and gravity casting. Furthermore, the mechanism of Si addition in Al-Cu alloys resulting in the dispersoids strengthening were studied using TEM analysis.

## II. Materials and methods

The experimental alloys with different Si contents were produced by melting pure Al (99.5%) and master alloys of Al-50% Cu, Al-10% Mn, Al-20% Si and Al-5% Fe (from Sichuan lande high-tech industry company, China). According to the Ref. 21, the optimum Mn addition is 0.6% to modify the Fe-rich intermetallic phase into less harmful shape. Fe is the common impurity in Al-Cu alloys and our aim focused on the high Fe impurity recycled aluminum alloys. Thus, the Fe content is 0.7 wt.% in present work. Different level of Si content: low Si-containing (Si content: 0 and 0.15wt.%), middle Si-containing (Si content: 0.55wt.%) and high Si-containing (Si content: 1.1wt.%) were used. Thus, the design alloy composition is Al-5.0Cu-0.6Mn-0.7Fe-XSi (X= 0, 0.15, 0.55 and 1.1%). The real chemical composition was analyzed by optical emission spectrometer (Perkin-Elmer Corporation, Optima 3000). The actual compositional of different alloys are Al-5.19Cu-0.64Mn-0.73Fe-0.03Si (alloy 1), Al-5.30Cu-0.63Mn-0.73Fe-0.15Si (alloy 2), Al-5.23Cu-0.63Mn-0.73Fe-0.55Si (alloy 3), Al-5.05Cu-0.67Mn-0.65Fe-1.12Si (alloy 4), respectively. Firstly, 10 Kg raw materials were melted at about 730 °C in a clay-graphite crucible using an electric resistance furnace and the melts were degassed by 0.5% $C_2Cl_6$ to minimize hydrogen content. The pouring temperature was set at 710 °C after degassing and the die was preheated to approximately 200 °C before squeeze casting. After the melt was poured into a cylindrical die, varied pressure (0 and 75 MPa) was applied to melts and held for 30 s until the melt was completely solidified. Finally, the ingots with a size of Φ 65 mm × 68 mm were obtained. The samples for T5 heat treatment were performed by solution heated to 538 °C for 12 h and then water quench. The aging process is performed at 155 °C for 8 h and then air cooling.

The tensile samples with dimension of Φ10 mm × 65 mm were cut from the edge of the ingots. The tensile test was performed on a SANS CMT5105 standard testing machine with a strain rate of 1 mm/min. At least three samples were tested to obtain average value. Samples for metallographic observation were cut from the end of tensile specimens. Then the sample were grinding different silica paper (100, 1000, 2500 type) and polishing. The samples for metallographic observation were etched with 0.5% HF solution for 30 seconds. Samples for grain size measurement were examined in Leica optical microscope (OM) with polarized light after anodizing with a 4% $HBF_4$ solution for about 30 s at 20 V. The volume fractions of Fe-rich intermetallics and the area fractions of dispersoids and precipitation free zone (PFZ) was analyzed with image analysis software Image-pro Plus. The measured area fractions of Fe-rich intermetallic phases were transferred as the volume fractions based on the assumption that the morphology of the Fe-rich intermetallic phases is equaled. Nearly 50 different fields were examined for each sample. The average chemical compositions of phases and fracture surfaces of tensile specimens were analyzed using scanning electron microscope (SEM) (Quanta 200), energy-dispersive X-ray analyzer (EDX). In particular, the morphology of the Fe-rich intermetallic phases and dispersoids were further studied by Tecnai G2 F30 field emission gun high resolution transmission electron microscopy (TEM) with energy dispersive X-ray (EDX) analysis.

**III. Results and discussion**

**A. Microstructure**

Fig. 1 shows the as-cast and heat treatment microstructure of the Al-5.0Cu-0.6Mn-0.7Fe alloys with different Si contents and applied pressure. The as-cast alloy usually consisted of α-Al matrix, Fe-rich intermetallic phases and $Al_2Cu$. It is seen that the black $Al_6$(FeMn), deep grey α-Fe and light grey $Al_2Cu$ in the matrix in the as-cast alloy 1, as shown in Fig. 1a. Fig. 1b shows the deep grey α-Fe and light grey $Al_2Cu$ existed in the in the as-cast alloy 4, which indicate that the size and amount of α-Fe increased with the increase of the content of Si. The heat-treated alloys consisted of α-Al matrix and Fe-rich intermetallic phases and the $Al_2Cu$ dissolved into the α-Al matrix during the T5 heat treatment, as shown in Fig. 1c-f. The results show the results that Si addition promote the transformation from α(CuFe) (the composition is the same as the as-cast state β-$Al_7Cu_2Fe$) to α-Fe at the heat treatment state. α(CuFe) and $Al_6$(FeMn) are the dominant Fe-rich intermetallic at low Si contents was shown in Fig. 1c. With the Si content increased to 0.15%, the $Al_6$(FeMn) gradually transformed into α-Fe (Fig. 1d), while a small volume percent of fine $Al_6$(FeMn) remained. This results is also consistent with previous report [21] that Si addition promote the formation of α-Fe due to the substituting of Al atom by Si atom in Fe-rich intermetallic phases. After the addition of 0.55% Si, the $Al_6$(FeMn) and α-Fe are existed in alloy 3, while the amount of α-Fe shapely increased and a small amount of $Al_6$(FeMn) (Fig. 1e). Further increasing the addition level of Si to 1.1%, α-Fe was the only Fe-rich intermetallic phase in the alloy 4 are shown in Fig. 1f. The effect of applied pressure on the Fe-rich intermetallic phases of alloys is shown in Fig. 1g and h. Compared to the alloy 1 without applied pressure (Fig. 1c), the size of α(CuFe) and $Al_6$(FeMn) become relatively smaller (Fig. 1g). Similarly, the size of α-Fe in the alloy with 75 MPa applied pressure became smaller and the branches of α-Fe become disconnected and closed to sphere shape (Fig. 1h). This indicating that applied pressure could be relatively refined the Fe-rich intermetallic phases. The volume percent of α-Al, α(CuFe), $Al_6$(FeMn) and α-Fe with different Si content and applied pressure has been measured as shown in Fig. 2. This further indicating that Si addition promotes the formation of α-Fe in heat treatment state. Moreover, the total volume percent of Fe-rich intermetallic phases under 75 MPa applied pressure are relatively smaller than those of without applied pressure. This is similar to the Ref. [20, 21], because more Cu and Mn atom solubility in α-Al matrix under applied pressure. The composition of Fe-rich intermetallic phases in as-cast and heat treatment state is given in Table I. It can be seen from the Cu content in heat-treated α(CuFe) is much higher those of in the as-cast $Al_3$(FeMn) and $Al_6$(FeMn) and Si, Cu and Mn contents in heat-treated alloys is relatively higher than those of in as-cast alloys. This indicating that Fe-rich intermetallic phases experiences solid-state transformation during heat treatment.

Fig. 3 shows the effect of applied pressure on the grain size of alloys and 3D morphology of Fe-rich intermetallic phases. Fig. 3a and b shows the grain size distributions of the alloy 1 with and without applied pressure, respectively. It can be seen that the grain size alloy with 75 MPa applied pressure (~300 µm) is much smaller than that of the alloy without applied pressure (~510 µm). The grain refinement effect is mainly attributed to the increase of melting point of alloy and the heat-transfer rates between the casting and die interface by eliminating air gaps [24]. Fig. 3c-f presents the deep-etched images of heat-treated Fe-rich intermetallic phases in alloys with different Si content and applied pressure. It can be observed that the α(CuFe) phase in alloy 1 have a cylindrical shape, as indicated in Fig. 3c. The convoluted branched structure α-Fe in alloy 4, which is coupled eutectic product grow from t00he large convoluted arm structure and α-Al dendrite [35], as shown in Fig. 3d. Compared with Fig. 3c and d, the size of Fe-rich intermetallic phases (Fig. 3e and f) in the alloy produced under 75 MPa applied pressure becomes comparatively smaller. This results are consistent with the microstructure features (Fig. 1). Thus, the applied pressure can not only refined the α-Al but also the Fe-rich intermetallic phases.

Fig. 4 shows the TEM analysis of Fe-rich intermetallic phases in the alloy 1 and 4. From the selected area diffraction pattern (SADP) in Fig. 4a, the A phases is identified as $Al_6(FeMn)$, which has an orthorhombic crystal structure with a lattice constant of a = 0.643 nm, b = 0.746 nm, c = 0.878 nm. The SADP result of B phase also confirms the crystal structure of α(CuFe), which has an tetragonal structure with lattice constant of a = b = 0.634 nm, c = 1.488 nm. The chemical composition of α(CuFe) is close to the as-cast state phase β-Fe. Fig. 4b and c shows the TEM images of the α-Fe in the alloy 4. Fig. 4b is a typical high-resolution TEM image, indicating the α-Fe (alloy 4) has the body-centered cubic (BCC) structure with lattice constant of a = 1.267 nm. This is further confirmed in the HRTEM image in Fig. 4c.

Besides the Fe-rich intermetallic phases, a dense distribution of T ($Al_{20}Cu_3Mn_2$) phase dispersoids in α-Al matrix is shown in Fig. 5a-d. It can be further confirmed by the indexed SADP and TEM micrographs as shown in Fig. 5e-h. It can be seen from Fig. 5a-d that the size of dispersoids decreased significantly while their quantity increased rapidly with increasing Si contents, especially for the alloys under applied pressure. A few precipitates can also be found in the intragranular regions (Fig. 5e). The interface between α-Al and T phase (as the red circle shown in the Fig. 5e) has been analysis by high-resolution electron microscopy (HREM) images, as shown in Fig. 5f. The precipitates and α-Al are further identified by selected area diffraction pattern (SADP), as shown in the Fig. 5f and g. The composition of T phase is Al: 79.21%, Mn:

12.17%, Cu: 7.40% and Fe: 1.21% (Fig. 5h). There are several possible explanations for the addition of Si content on enhancing the T precipitation. First, Si may provide nuclei for the heterogeneous nucleation and retards the coarsening of the precipitates. Because the presence of Si in the alloy reduces the solubility of Mn in α-Al matrix [36], thereby increasing the chemical diving force for T phase precipitation and reducing the α-Al/T interfacial free energy. Moreover, the aspect ratio of T phase decrease with the increase of Si content, which means that Si addition increase the coarsening resistance of precipitates T phases. The high number density of T phase precipitates is responsible for explain the small size and low aspect ratio. Si and precipitates in alloys have a strong elastic interaction because of their compensating strain fields, which promotes the nucleation of precipitates on Si, decreases the expected aspect ratio of precipitates, and inhibits coarsening. Also, Si addition increase attractive binding energy between Si and vacancy, which can act as heterogeneous nucleation sites for dispersoids T phases [37]. Moreover, high diffusivity of Si in α-Al matrix can accelerate the diffusion kinetics of elements Cu and Mn and reduces the T phase precipitate coarsening resistance [38].

The schematic image of the microstructure evolution in the alloys with different Si contents is shown in Fig. 6. After the alloys being solid-state treatment at 538 °C for 12 hours, the entire amount of $Al_2Cu$ is dissolved into the Al matrix, as shown in Fig. 1c-h. During the solid-state treatment, the Fe-rich intermetallic phases become unstable and fragmented and transformed into newly different Fe-rich intermetallic phases. Compared to the morphology of the as-cast Fe-rich intermetallic phases (Fig.1a-b), no significant change of the morphology of the Fe-rich intermetallic phases has occurred during the heat treatment except for some Fe-rich intermetallic phases become less connected. In low Si alloy (Alloy 1 and 2), α(CuFe) nucleates on the interface between α-Fe/Al$_6$(FeMn) and α-Al matrix (Fig. 4a) and the Cu atoms diffuse into the α-Fe/Al$_6$(FeMn) during the heat treatment, as shown in Fig. 6a-d. Because the high density of dislocations on the interface of α-Fe/Al$_6$(FeMn) and α-Al matrix. Once the α(CuFe) nucleation, it will quickly growth in a manner of dendritic growth. This is can be attributed to the relatively high diffusion rate of Cu and Mn at 538 °C. This results is similar to previous Ref. [21]. In addition, there is a small amount of Al$_6$(FeMn) remained in the alloy due to lack of sufficient Cu to form α(CuFe). Thus, the solid-state transformation of alloy 1 and 2 is a eutectoid reaction:

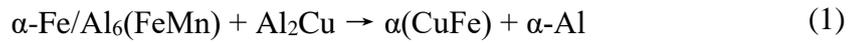

$$\alpha\text{-Fe/Al}_6(\text{FeMn}) + \text{Al}_2\text{Cu} \rightarrow \alpha(\text{CuFe}) + \alpha\text{-Al} \qquad (1)$$

In high Si containing alloys (Alloy 3), the dominant α-Fe and a small amount of α(CuFe) distribute in the matrix. While there is only α-Fe in the alloy 4, indicating the occurrence of eutectoid reaction with increasing Si content:

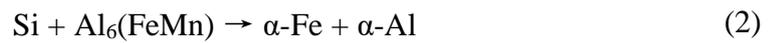

$$\text{Si} + \text{Al}_6(\text{FeMn}) \rightarrow \alpha\text{-Fe} + \alpha\text{-Al} \qquad (2)$$

α-Fe usually nucleates on the interface between the Al$_6$(FeMn) and α-Al matrix and consumption progressively the Al$_6$(FeMn) (Fig. 6e-h). The phase transformation from Al$_6$(FeMn) to α-Fe is called "6-to-α transformation", and Si is the key factor for this transformation [36-39]. The free Si atom diffusion into Al$_6$(FeMn) and decompose to a mixture of α-Fe and α-Al. The decomposition of Al$_6$(FeMn) perserves the volume and content of Fe and Mn, and the requirement the intake of Si. If more free Si atom in the surrounding α-Al matrix, the rate of "6-to-α transformation" increases. Because the diffusion coefficient of Si in α-Al at 500 °C is about $1.4 \times 10^{-13}$ $m^2s^{-1}$ [33], which is much higher than that of Cu and Mn. Si in the matrix diffusing into the $Al_6$(FeMn), which accelerate the eutectoid reaction. With the increasing of Si content, the volume percent of α-Fe increases in the alloy 3. For the alloy 4, the α-Fe keep thermodynamic stable because Si is the key element for the α-Fe [38, 39]. The increasement of Si atom in the α-Al matrix, the thermodynamic stable phase change from $Al_6$(FeMn) to α-Fe. Previous studies [39] also confirmed this, the nucleation of the α-Fe phase is the overall rate-controlling factor for the 6-to- α transformation. If a greater source of silicon could be supplied in the alloys, the growth of the α-phase-Al eutectoid through a particle is relatively fast. Moreover, Si promotes thermodynamic stable of α-Fe and prevents the transformation from α-Fe to α(CuFe) because Si can substitutes Al and Si can substitutes Cu in α-Fe.

**B. Mechanical properties**

The effect of Si content on the mechanical properties of heat-treated Al-5.0Cu-0.6Mn-0.7Fe alloys is shown in Fig. 7. There is a considerable increment in the ultimate tensile strength (UTS) and the yield strength (YS), but a significant decrement in the elongation in the alloys with increasing the Si contents from 0% to 1.1%. For example, Compare with the alloy 1 (0% Si), UTS and YS of alloy 4 (1.1% Si) without applied pressure were 29.7%, 101.5% higher than those of alloy 1. It can also be seen that the applied pressure improved the mechanical properties of the alloys. The UTS, YS and elongation of the squeeze cast Al-5.0Cu-0.6Mn-0.7Fe-1.1Si alloys is 386 MPa, 280

MPa and 8.6 %, respectively, which is about 21.2%, 6.9%, 73.9% higher than that of the alloy without applied pressure. In addition, the Si particles in the high Si content alloys (alloy 4) also benefit the improvement of strength. According to previous study [40], the small Si peak could be found in the XRD analysis. Although the Si particles could not be found in the SEM images, which maybe owing to the similar atom weight of Al and Si.

Table II summarizes the mechanical properties of Al-Cu alloys with various Fe and Si contents in literature [13, 16, 21, 41 and 42] and present study. These alloys are prepared by different technique, including gravity casting (GC), squeeze casting (SC) and high pressure die casting (HPDC), and their alloy composition focused on low Fe and Si impurities content. Ref. 21 reported the high impurity Fe content in the Al-5.0Cu-0.6Mn-1.0Fe alloy, while their tensile properties (UTS and elongation) is not very good. The UTS values of the Al-Cu alloys reported by the Refs. [13, 16] is relatively high, while their elongation values is relatively low. The HPDC processed Al-4.4Cu-0.2Fe-1.2Si-0.4Mg-0.2Ti alloys [42] possess excellent tensile properties (combined UTS and elongation), while their Fe impurity content is relatively low which are not suitable for the recycled alloys. It can be seen that the alloy 4 with high Fe and Si impurities shows combined high strength and ductility. In this regard, the squeeze casting technique can relatively extend the Fe and Si limitation in the Al alloys. Moreover, the mechanical properties of the alloy 4 meet the requirement of safety-critical suspension component in the automotive industry (UTS> 380 MPa, elongation> 7%). This means that the present study is helpful to promotion of high efficient utilization of recycled aluminum alloys and it is also available for reduce the manufacture cost.

In order to further evaluate the mechanical properties, the Quality Index (QI) was used to characterize the casting quality. The QI was firstly introduced by Cáceres *et al.* [43, 44] to address the quality index. This index is obtained according to the variation of UTS with elongation obtained with alloys submitted to different metal treatment, heat treating or alloys composition. And the QI of the studied alloys can be calculated by the following equation [43]:

$$QI = UTS + d\log El \tag{3}$$

Where QI stand for Quality Index, UTS and El stand for ultimate tensile strength and elongation respectively, $d$ is empirically determined constants (we take 270 for $d$ in the present study). It is found that the QI value of the alloy 4 without applied pressure is 522 MPa, while that of the alloy 4 with applied pressure of 75 MPa is 656 MPa. Due to the higher UTS and elongation, the QI value of the alloy 4 increases with increasing applied pressure. In addition, the QI value varies with different chemical composition, such as Si contents. It is found that the QI value of alloy 1 without applied pressure is 531 MPa, while 522 MPa for the alloy 4. Nevertheless, the QI value of alloys varies with increasing Si content at the same level of applied pressures. Overall, the applied pressure is helpful to improve the QI and increasing Si content keep QI value at the similar level. Consequently, the Si content limitation can be relatively relax under applied pressure. The best QI values of 656 MPa obtained in this study is the alloy 4 with applied pressure of 75 MPa. While, Kamga *et al.* [16] reported the B206 (one kind of typical Al-Cu-Mn) alloys with the low Fe and Si content (0.1 - 0.3%) under T4 and T7 heat treatment. They found that when the Fe/Si ratio close to one and high cooling rate, the high mechanical properties are obtained. Their QI value range of 557 - 640 MPa for T7 heat treatment and their QI value range of 727 - 810 MPa for T4 heat treatment. It can been see that the natural aging (T4 heat treatment) can reaches higher QI value because of the higher elongation. The reported literature [16] has a higher QI value than the present study, because of the low Fe and Si content in the B206 alloys (up to 0.3%). Thus, in order to obtain the good casting quality and high QI value, we should take the alloy composition, heat treatment and cooling rate into consideration.

The above tensile testing shows that Si addition enhances the strength of the alloys. The main reason is that the volume percent of α-Fe increases and α(CuFe) decreases (Fig. 2) and dispersoid strengthening of T phases with increasing Si content (Fig. 5). The α-Fe is less harmful than α(CuFe) because the Chinese script α-Fe has a compact structure, which make the cracks more difficult to initiate and propagation during tensile test. Usually, α-Fe have well-developed branches in different directions, they mutually interwoven with the Al matrix [13]. During the tensile test, the refined α-Fe branches block the crack propagation and it only can propagation along the α-Fe/matrix interface. Also, the void cannot easily developed around the α-Fe. The other factor mainly attribute to the dispersoid strengthening in the matrix. The amount of T phase in the matrix increases and their size decreases with increasing Si content during T5 heat treatment. The yield strength of the alloys ($\sigma$) is usually composed of three parts:

the strength of dispersoid strengthening ($\sigma_D$); the strength of solid solution strengthening ($\sigma_{SS}$) and the strength of α-Al matrix ($\sigma_{Al}$) [45].

$$\sigma = \sigma_D + \sigma_{SS} + \sigma_{Al} \quad (4)$$

In the present work, the T phase dispersoid strengthening in the matrix and solid solution strengthening both contribute the improvement of the strength. For example, with the Si content increase from 0% to 1.1% without applied pressure (Table III and Fig.5a and c), the incensement of strength is 12.98 MPa. This is can be attributed to solid solution strengthening of Si addition. A large numbers of fine precipitates in the matrix pose a hindrance to moving dislocations resulting in the enhancement of strength. And the strength of α-Al matrix is 5.14 MPa. The solid solution strengthening is mainly depend on the concentration of solute in the solid solution. While, the Fe, Si, Mn and Cu solute atoms in Al-Cu alloys dissolve into the matrix contribution to the improvement of the strength. Because of the low solubility of Fe and Si in the matrix, their solid solution strengthening contribution to the improvement of strengthen can be neglected. According to the Ref. 41, the solid solution of Cu and Mn in Al-5.0Cu-0.6Mn-1.0Fe is increased with the improvement of applied pressure. For alloy 1, with the applied pressure increase from 0 MPa to 75 MPa (Table III and Fig.5a and b), the incensement of strength is 2.06 MPa. Because the size of dispersoids T phase is relatively large, the Orowan mechanism can be applied in the present study. The contribution of dispersoid strengthening can be calculated by the Ashby-Orowan equation [45, 46]:

$$\sigma_D = \frac{0.84MGb}{2\pi(1-\nu)^{1/2}\lambda} \ln \frac{r}{b} \quad (5)$$

where M is the Taylor factor, M = 3.06; G is shear modulus of the Al matrix, G = 27.4 GPa; b is the Burgers vector of dislocation in Al, b = 0.286 nm; ν is the Poisson ratio, for Al, ν = 0.33. The interspacing of dispersoids λ depends on the radius r and volume fraction f of dispersoids [47]:

$$\lambda = r\left(\frac{2\pi}{3f}\right)^{1/2} \quad (6)$$

The parameter for calculation the dispersoid strengthening of the alloys and the calculated dispersoid strengthening $\sigma_D$ are list in the Table III. In the alloy with 0% and 1.1% Si content without applied pressure, the equivalent diameter of the T phase is 1.07 μm and 0.33 μm, respectively; and their corresponding number density is about $5.75 \times 10^{20}$ m$^{-3}$ and $48 \times 10^{20}$ m$^{-3}$, respectively; their corresponding dispersoid strengthening is about 5.14 MPa and 18.12 MPa, respectively. It can be found that the equivalent diameter of the T phase decreases by about 69%, the number density and dispersoid strengthening increases by at least approximately 734% and 253%, respectively. Thus, it can be deduced that the addition of Si have a great influence on the precipitation of T phase. Moreover, an increase in the applied pressure from 0 to 75 MPa provides precipitation of the T phase with a smaller equivalent diameter and higher number density. For example, the equivalent diameter decreased by nearly 9%, the number density and volume fraction of alloy 1 increases by approximately 97% and 40%, respectively, which fully confirms the applied pressure also play a major role in the precipitation of T phase. These indicating that Si addition and applied pressure both promote the formation of high density small dispersoids T phase, and contribution to the enhancement of yield strength.

The decrease of elongation with Si addition can be concluded from fracture characterization. Fig. 8 shows the SEM images of fracture morphology and longitudinal sections near the fracture surfaces of the alloys. It is clear from low Si alloys (Fig. 8a and b) that large numbers dimples exist at fracture surface. While, many coarse cleavages and microcrack are observed in medium and high Si alloys (Fig. 8c and d). These indicating that increasing Si contents change the fracture morphology from ductile to quasi-cleavage. Moreover, the applied pressure is helpful to elimination porosity. The previous study demonstrated that the detrimental effect of compact α-Fe is less than α(CuFe) [13]. However, the increasing volume of α-Fe in the alloys with increasing Si content still resulting in the decreasing of elongation. The longitudinal sections near the fracture surfaces of different alloys are shown in Fig. 8e-h. It can be seen that large numbers of porosities are observed at the fracture surface in alloys without applied pressure (Fig. 8e and g), which is the site for crack initiation. The crack are usually exist in coarse Fe-rich intermetallic phases (Fig. 8f and h) indicating that the applied pressure is helpful to eliminate the porosities and hinder the crack propagation.

## IV. Conclusion

The microstructure and mechanical properties of the T5 heat-treated Al-5.0Cu-0.6Mn-0.7Fe alloy with different Si contents produced by gravity casting and squeeze casting have been investigated to understand the role of Si on the Fe-rich interetallic phase formation and their effect on the tensile properties. The main conclusions are as follows:

(1) The increasing Si content increases the volume fraction of Fe-rich intermetallic phases and promotes the formation of α-Fe. With addition 0.55% Si in Al-5.0Cu-0.6Mn-0.7Fe alloy, α-Fe formed through the eutectoid reaction: Si + $Al_6$(FeMn) → α-Fe + α-Al; with further increased the Si content to 1.1%, the α-Fe keep thermodynamic stable.

(2) The addition of Si in Al-5.0Cu-0.6Mn-0.7Fe alloy enhances the UTS and YS due to the increasing volume fraction of T phase and the less harmful α-Fe has a compact structure, which make the cracks more difficult to initiate and propagation during tensile test. The increase of Si content resulting in fine high dense T phases, the dispersoid strengthening contribution the improvement of tensile properties.

(3) The Al-5.0Cu-0.6Mn-0.7Fe alloy with 1.1% Si addition under 75MP applied pressure shows the best mechanical properties, which were UTS: 386 MPa, YS: 280 MPa and elongation 8.6 %, respectively.

**Acknowledgement**

The authors would like to acknowledge the financial support from Project (51374110) by the National Natural Science Foundation of China and Project (2015A030312003) by the Natural Science Foundation of Guangdong Province for Research Team. Yuliang Zhao also would like to acknowledge the financial support from Chinese Scholarship Council (CSC).

**Figure Captions**

Fig. 1. Microstructure of the Al-Cu-Mn alloys with different Si content and applied pressure: (a) as-cast alloy 1 without applied pressure; (b) as-cast alloy 4 without applied pressure; (c) heat-treated alloy 1 without applied pressure; (d) heat-treated alloy 2 without applied pressure; (e) heat-treated alloy 3 without applied pressure; (f) heat-treated alloy 4 without applied pressure; (g) heat-treated alloy 1 with 75 MPa applied pressure; (h) heat-treated alloy 4 with 75 MPa applied pressure.

Fig. 2. The volume percentage of different phases in the heat-treated alloys under different applied pressures.

Fig. 3. Effect of applied pressure on the grain size: (a) alloy 1 without applied pressure and (b) alloy 1 with 75 MPa applied pressure; 3D morphology of Fe-rich intermetallics of heat-treated alloys: (c) heat-treated alloy 1 without applied pressure; (d) heat-treated alloy 4 without applied pressure; (e) heat-treated alloy 1 with 75 MPa applied pressure; (f) heat-treated alloy 4 with 75 MPa applied pressure.

Fig. 4. TEM analysis of the Fe-rich intermetallics: (a) bright-field image of $Al_6(FeMn)$ and α(CuFe) and corresponding SAED pattern in alloy 1; (b) TEM image of α-Fe and corresponding SAED pattern in the alloy 4: (c) the HRTEM image of α-Fe.

Fig. 5. TEM images showing the morphology and density of T phase dispersoids in heat-treated alloys: (a) alloy 1 without applied pressure; (b) alloy 1 with 75MPa applied pressure; (c) alloy 4 without applied pressure; (d) alloy 4 with 75MPa applied pressure; (e) T phase in the matrix of alloy 1; (f, g) selected area diffraction pattern (SADP) of T

phase/matrix interfaces and their schematic diagram; (h) the chemical composition of T phase.

Fig. 6. Schematics of the microstructure evolution of alloys with different Si contents: (a-d) low Si content alloys; (e-h) high Si content alloys; (a, e) the as-cast alloy; (b, f) solution treatment early stage: the dissolution of $Al_2Cu$ phase; (c, g) solution treatment late stage: the fragmentation of Fe-rich intermetallic phases; (d, h) aging: precipitation fine T phases.

Fig. 7. Effect of Si content on the mechanical properties of heat-treated Al-5.0Cu-0.6Mn-0.7Fe alloys: (a) UTS, (b) YS, and (c) elongation.

Fig. 8. Fracture surface of heat-treated Al-5.0Cu-0.6Mn alloys with different Si contents and applied pressures: (a) alloy 1 without applied pressure; (b) alloy 1 with 75 MPa applied pressure; (c) alloy 4 without applied pressure; (d) alloy 4 with 75 MPa applied pressure; longitudinal sections near the fracture surfaces: (e) alloy 1 without applied pressure; (f) alloy 1 with 75 MPa applied pressure; (g) alloy 4 without applied pressure; (h) alloy 4 with 75 MPa applied pressure.

**Tables**

Table I Chemical composition of the Fe-rich intermetallics in the as-cast and heat-treated conditions (at. %).

| Conditions | Alloys | Phase | Al | Cu | Mn | Fe | Si |
|---|---|---|---|---|---|---|---|
| As-cast | 1 | $Al_6(FeMn)$ | 80.05±1.78 | 5.12±0.85 | 2.64±0.34 | 12.19±1.91 | - |
|  | 4 | α-Fe | 73.45±0.85 | 1.98±0.35 | 3.02±0.60 | 9.89±0.24 | 8.90±0.55 |
| Heat-treated | 1 | $Al_6(FeMn)$ | 85.36±1.67 | 3.08±0.31 | 2.76±0.22 | 8.79±1.21 | - |
|  | 1 | α(CuFe) | 61.95±1.77 | 22.10±1.08 | 2.53±0.49 | 13.41±0.89 | - |
|  | 4 | α-Fe | 71.69±1.93 | 3.71±0.28 | 5.04±0.49 | 9.54±0.90 | 10.12±0.77 |

Table II Mechanical properties of Al-Cu alloys in literature and present study.

| Alloy | Processing* | Heat treatment** | UTS (MPa) | Elongation (%) | Reference |
|---|---|---|---|---|---|
| Al-5.0Cu-0.6Mn-1.0Fe | GC | T5 | 250 | 5.5 | [41] |
| Al-5.0Cu-0.6Mn-1.0Fe | SC | T5 | 290 | 14 | [41] |
| Al-4.6Cu-0.3Mn-0.5Fe-0.3Mg-0.3Si-0.2Ti | GC | T7 | 424 | 2.8 | [13] |
| Al-4.7Cu-0.2Mn-0.3Fe-0.3Si-0.3Mg | GC | T7 | 510 | 1.5 | [16] |
| Al-4.4Cu-0.2Fe-1.2Si-0.4Mg-0.2Ti-1.2Si | HPDC | T7 | 370 | 9 | [42] |
| Al-5.0Cu-0.7Fe-0.6Mn-1.1Si | SC | T5 | 386 | 8.6 | Present study |

* GC - gravity casting, SC - squeeze casting, HPDC - high pressure die casting.

** Heat treatment: T4: solution treatment at 505 °C / 2h + 520 °C /8h + aging at room temperature/8d; T5: solid state: 538 °C 12 h + aging: 155 °C 8 h; T7:solid state: 505 °C 2h + 520 °C 8h + aging: 185 °C 5h.

Table III The parameter for calculation the dispersoid strengthening of the alloys.

| Alloys | Applied pressure (MPa) | Equivalent diameter D (μm) | Volume fraction f (%) | Number density N ($10^{20}$ m$^{-3}$) | Volume fraction of dispersoids zone (%) | Volume fraction of dispersoids free zone (%) | $\sigma_D$ (MPa) |
|---|---|---|---|---|---|---|---|
| 1 | 0 | 1.07 | 1.76 | 5.75 | 2.71 | 19.66 | 5.14 |
| 1 | 75 | 0.97 | 1.97 | 11.33 | 3.33 | 25.37 | 7.20 |
| 4 | 0 | 0.33 | 2.94 | 48 | 7.92 | 34.69 | 18.12 |
| 4 | 75 | 0.23 | 2.49 | 69 | 9.08 | 42.65 | 23.80 |